# Explanatory Publics: Explainability and Democratic Thought

David M. Berry

In order to legitimate and defend democratic politics under conditions of computational capital, my aim is to contribute a notion of what I am calling *explanatory publics.* I will explore what is at stake when we question the social and political effects of the disruptive technologies, networks and values that are hidden within the "black boxes" of computational systems. By "explanatory publics", I am gesturing to the need for frameworks of knowledge — whether social, political, technical, economic, or cultural — to be justified through a *social right to explanation.* That is, for a polity to be considered democratic, it must ensure that its citizens are able to develop a capacity for explanatory thought (in addition to other capacities), and, thereby, able to question ideas, practices, and institutions in society. This is to extend the notion of a public sphere where citizens are able to question ideas, practices, and institutions in society more generally.[1] But it also adds the corollary that citizens can demand explanatory accounts from institutions and, crucially, the digital technologies that they use. I agree with Outhwaite that "what makes an explanation, or what makes an explanation a good one, is therefore a difficult question, which may require a detailed study, not just of the logical properties of the explanation but of the context in which it was offered".[2]

This is important in computational capitalism because when we call for an *explanation,* we are able to understand the contradictions within this historically specific form of computation that emerges in late capitalism. These contradictions are continually suppressed in computational societies, but generate systemic problems borne of the need for the political economy of software to be obscured, so that its functions and the mechanisms of value generation are hidden from public knowledge. Why should this fundamental computational political economy be concealed? One reason is that an information society requires a form of public justification in order to legitimate it as an accumulation regime and to maintain trust. Trust is a fundamental basis of any system and has to be stabilised through the generation of norms and practices that create justifications for the way things are. This is required, in part, because computation is increasingly a central aspect of a nation's economy, real or imaginary. The suppression of its political economy is required, because computation rapidly destabilises the moral economy of capitalism, creating vast profits from exchange and production processes that might be considered pre-capitalistic or obscenely inegalitarian such as intensive micro-work or fragmented labour in the gig economy. These more recent experiments with micro-task production are nothing less than attempts to reinvent the world as a post-factory society. It requires the building of a new infrastructure of production by enclosing labour-power within algorithmic "wrappers" that present the surface effect of a seemingly unending stream of abstract labour. This labour-power is made available via websites and apps, creating a highly alienated form of labour-power that is disciplined and managed algorithmically through various forms of "signal" mechanisms that are generated by the system such as pay, ratings, reviews, and metrics. The "boss" of the old factory is abolished by computation and replaced by the algorithm that guides, chides, and informs through a personal device such as a smartphone, whose very intimacy makes it compelling and trustworthy.

Further, many of the sectors affected by computation are increasingly predicated on the illegitimate manipulation or monopolisation of markets, or are heavily data extractive. These effects threaten individual liberty, undermining a sense of individual autonomy, and even that bulwark of the neoliberal system: consumer sovereignty. Profit from computation also often appears to require the mobilisation of persuasive technologies that cynically, but very successfully, manipulate addictive human behaviour. Therefore, one of the key questions we need to ask is: How much computation can a society withstand? We can only answer this question if we create new forms of explanatory publics that have competences to discuss, critique, and challenge computational technical systems.

Google, at least at one point, internally understood the problem of an excess of computational power in terms of what it called a "creepy line".[3] Within the line, public acceptance of computation generates huge profits (or "good computation"), and outside of which computation is able to create effects that would be politically, or economically problematic, or even socially destructive, but which might generate even larger profits (or "bad computation"). The founders of Google, Larry Page and Sergey Brin, gestured towards this in their famous paper from 1998, "The Anatomy of a Large-Scale Hypertextual Web Search Engine", where they warned that if the "search engine were ever to leave the 'academic realm' and become a business, it would be corrupted. It would become 'a black art' and 'be advertising oriented'".[4] As Carr describes,

> That's exactly what happened — not just to Google but to the internet as a whole. The white-robed wizards of Silicon Valley now ply the black arts of algorithmic witchcraft for power and money. They wanted most of all to be Gandalf, but they became Saruman.[5]

Peter Thiel, a PayPal co-founder and chairman of Palantir, revealed a similar tendency made possible by computation when he identified the importance of software companies securing a technical monopoly. He termed this as a movement from "zero to one", the "one" representing the successful monopolisation of a technical niche or sector of the economy.[6] While this is not necessarily a surprise, the candour with which the Silicon Valley elite advocate for these economic structures, which are contrary to neoliberalism, let alone social democracy, should give us pause for thought. Indeed, Thiel goes so far as to argue that he "no longer believe[s] that freedom and democracy are compatible".[7] But while the profit-oriented organisation of a capitalist economy is unchanged, what is new is that exploitative processes function at a new intensity, and at all levels of society due to computation. It is no longer just the workers who are subject to processes of automation, but also the owners of capital themselves and, inevitably, their private lives. That the millionaires and billionaires of the technology industry should feel a need to protect their own families from the worst aspects of computation, with Steve Jobs famously withholding computers from his children and Larry Page, one of the co-founders of Google, managing to keep his personal life and even his children's name secret, is ironic given Google's mission "to organise the world's information and make it universally accessible and useful".[8] Unfortunately, this disconnectionism is not an option available to the majority of the world's population — even as it becomes a bourgeoise aspiration through digital detox camps and how-to-disconnect-guides in national newspapers.

The contradictions generated by this new system can be observed in discourse. Concepts carry over from the computational industries and spread as explanatory ideas across society. Indeed, we see principles from software engineering offered up for social engineering, with open source identified as an exemplar principle of organisation; platforms as future models for governance; calculation substituted for thought; and social media networks replacing community. This can also be seen when computation is described through dichotomies such as transparent and opaque, open and closed, augmentation and automation, freedom and subjugation, resistance and hegemonic power, the future of the economy, and its destruction.

If we focus on two of these discursive categories, augmentation and automation, we can see how they are used to orient and justify further computation. As far back as 1981, Steve Jobs, then CEO of Apple, famously called computers "Bicycles for the Mind", implying that they *augmented* the cognitive capacities of the user, making them faster, sharper, and more knowledgeable. He argued that when humans "created the bicycle, [they] created a tool that amplified an inherent ability [...] The Apple personal computer is a 21st century bicycle if you will, because it's a tool that can amplify a certain part of our inherent intelligence [...] [It] can distribute intelligence to where it's needed".[9] This vision has been extremely compelling for technologists and their apologists, who omitted to explain that these capacities might be reliant on wide-scale surveillance technology. But whilst this vision of bicycles for the mind might have been true in the 1980s, changes in the subsequent political economy of our societies means that computers are increasingly no longer augmenting our abilities, but rather *automating* them. Algorithms then become Weberian "iron cages", in which citizens are trapped and monitored by software, with code that executes faster than humans can think, overtaking their capacity for thought. Augmentation, which extends our capacity to do things, and automation, which replaces this capacity, show how analysing the historically specific examples of key dyadic concepts is crucial for understanding this struggle over the future of computation. Indeed, in response, we need a way of transcending these dichotomies, linking computation to a call for a social right to explanation — through what I call *explanatory* publics.

One of the ways in which we can do this is by recognising how information economies are founded on an attempt to make *thought subject to property rights.*[10] Principles of reasoning, mathematical calculation, logical operations, and formal principles necessarily become owned and controlled for an informational economy to function. But these forms of thought also become recast as the only legitimate forms of reason, feeding back into a new image of thought. Data is increasingly associated with wealth and power, linked explicitly with computational resources that submit this data to rapid computation and pattern-matching algorithms through machine-learning and related techniques. Humans can now purchase thinking capacity, whether through special algorithms or the augmentation possibilities of personal devices. Information processing is now so fast that it can be performed in the blink of an eye, and the results used to augment, if you can afford it, or else persuade, and potentially manipulate others who cannot. Depending on the price you're willing to pay, digital corporations can sell algorithms to either increase, or undermine an individual's reasoning capacities, and thereby supplement or substitute artificial analytic capacities that bypass the function of reason. For the wealthier, they have the option to literally buy better algorithms, better technologies, better capacities for thought. For the rest of us, algorithms overtake human cognitive faculties by shortcutting individual decisions by making a digital "suggestion" or "nudge". Indeed, many technology companies rely on techniques developed in casinos to nudge behaviour to maximise profitability such as creating addictive experiences and by disarming the will of the user.[11] As a consequence, cognitive inequality emerges in relation to a new neuro-diversity created by augmenting, or automating thought itself, potentially undermining democratic and public values. An example of how computation can differently segment the market is presented by the Amazon Kindle, which comes in two varieties: a cheaper version that contains constantly updating advertising on its home and lock screen ("With Special Offers"), and a more expensive version that is free of adverts ("Without Special Offers").

We might also note that the actually existing informational economy is built increasingly on software that has steered capitalism towards a data-intensive form of extractive economy — what Zuboff has termed "surveillance capitalism", and Stiegler has identified as "the automatic society".[12] This has been achieved through spying on users, data capture, arbitrage, and the manipulation of markets, but also, crucially, through facilitat-ing monopolies of data — by using digital rights management, copyright, or patents. One of the scandals of contemporary capitalism is the extent to which widescale data capture and monitoring of users, their private lives, and their economic activities has been facilitated by computation. Not only has this been relatively unregulated, but it has allowed companies to assume that this kind of wholesale spying on people is the new normal and an acceptable practice in business. The scale of automated data accumulation is completely without precedent historically. Just taking Facebook as an example, we see almost continuous data collection on over 1.2 billion people worldwide. So much data, in fact, that even the CIA, the US intelligence agency, has signalled its inability to deal with the overload from what they call "digital breadcrumbs".[13] Indeed, Ira "Gus" Hunt, the CIA's former chief technology officer, has argued that "the value of any piece of information is only known when you can connect it with something else that arrives at a future point in time [...] Since you can't connect dots you don't have [...] we fundamentally try to collect everything and hang on to it forever".[14]

This has led to the idea that perhaps data contains highly lucrative insights that can create new sources of profit. While discourses about how "data is the new oil" have circulated, inevitably, those companies keen not to miss a profitable opportunity have put aside their caution and defaulted to maximum data collection whenever and wherever possible. Opting out of this surveillance regime has also become progressively more difficult, and, in my own case, attempting to "opt out" of the numerous data collection companies associated with the *Huffington Post,* for example, took three hours of frustrating clicking through numerous privacy statements. Even then, key links and options would be disguised as "hidden links", sliders that "were not available at the moment" and other techniques of dissuasion.[15] These companies build these systems deliberately in a user-hostile way, while keeping the default of data collection, even of minors and others who cannot legally give consent, in a system that, nonetheless, is structured in such a way as to unequally distribute the effects of data collection and algorithmic profiling to the poorer, less well-educated segments of society.

These data monopolies signal fundamental contradictions at the heart of computational capital, which appears to require legitimation through a façade of progress, individual choice and an enlightened technical philosophy, while the actually existing underlying political economy is increasingly structured around distortion, deception, and data capture.[16] These conditions create a form of cynical reason, particularly in the software industry, that I term *neo-computationalism.* This is an ideology subscribed to by an unhappy consciousness, which disavows the use of unsavoury data collection and surveillance techniques, while continuing to practice it. It is a system of thought that holds to a belief that social problems can be solved using more computation, at the same time as creating technical systems and algorithms that make them worse or amplify their pathologies. This contradiction at the level of both political economy and individual consciousness is destabilising to society and cannot be kept in check without the mobilisation of a set of justificatory discourses through the ideology of neo-computationalism.

Peter Sloterdijk has described cynical reason as enlightened false consciousness, which "is afflicted with the compulsion to put up with preestablished relations that it finds dubious, to accommodate itself to them, and finally even to carry out their business". Sloterdijk quotes Gottfried Benn, who explains modern cynicism as that which is lived as a private disposition that requires you,

> to be intelligent and still perform one's work, that is
> unhappy consciousness in its modernized form, afflicted
> with enlightenment. Such consciousness cannot become
> dumb and trust again; innocence cannot be regained. It
> persists in its belief in the gravitational pull of the relations
> to which it is bound by its instinct for self-preservation. In for
> a penny, in for a pound. At two thousand marks net a month,
> counterenlightenment quietly begins; it banks on the fact
> that all those who have something to lose come to terms
> privately with their unhappy consciousness or cover it over
> with "engagements".[17]

Neo-computationalism extols an epistemology of computation that fetishises the surface — that refers to knowledge in, and through the interface of a computer. This surface, which may be represented visually, aurally, or through haptics, becomes accepted *as* the computational. For example, one of the most seductive representations of computation has become the "network". This is often represented visually through points and lines connected together in a highly distributed manner. That is not to say that networks don't have an important place in the technical infrastructure — clearly they do, as this network model is fundamental to the design of the Internet. However, the network should not be seen as an ontology, it cannot and does not explain everything about computational systems. Indeed, it can hide more than it reveals as an explanatory framework.

We might, therefore, understand the network as an "apparatus of the dark" comparable to the lightning that Emily Dickinson memorably described as generating ignorance of what lies behind in "mansions never quite revealed".[18] In response to the poverty of the network, there have been serious attempts to understand the fundamental mechanisms of computation through a turn to stacks, infrastructure, materiality, code, software, and algorithms to try to uncover aspects of the computational that have been hidden, or that are difficult to discern. However, I argue that under neo-computationalism, the illegibility of the information society's systems is seen as necessary for it to function and must be generally accepted as a *doxa* of modern society — even as a desirable outcome. If we do not have to see the ugliness of the underlying logics of computational capitalism, then one does not have to come to terms with it, we can merely ignore it, disguised as it is behind the post-digital interfaces of our modern smartphones and laptops.

This logic of obscurity has justified the proprietary economic structure of software intellectual property rights through a technical division between source-code and execution, and the principles of object-oriented design, in which the mechanisms of computation are kept obscured or hidden. I argue that these two aspects of knowing computation — surface and mechanism — are a result of this underlying political economy, which generates a fundamental bifurcation in knowledge in computational societies.

This division of knowledge between a seen and hidden realm is often justified through concepts of simplicity, ease of use, or as convenience — most notably, by the technology industries, especially the so-called FAANG companies (Facebook, Apple, Amazon, Netflix, Google). I argue that one of the outcomes of this is the turn to "smartness" as a justificatory discourse through "operational functionality"; namely, that "smart" results justify the opacity of the hidden aspect of this epistemology. Smartness and opacity are, therefore, directly linked through an epistemological framework that establishes a causal link between data and "truth", but not through a veracity that requires the material links in the chain of computation to be enumerated or understood. In other words, ignorance of computational processes is, under this epistemology, celebrated as a means to an end of smartness. One of the results is to locate data as the foundation of computational inequities or computational power. Injustice is strongly linked to data problems, which, some have argued, can be addressed by more data, ethical data, or democratising data sources. The recent explosion of literature on data ethics and the eagerness with which the technology industry has taken it up, might be explained by the weakness of its critical edge. Google and Facebook have both set up "ethics" groups, often staffed by academic ethicists, although Google promptly had to dissolve its committee after an outcry over its membership.[19] Ultimately though, data ethics has proven to be unsurprisingly toothless when confronted with the forces of data collection and surveillance, and more adept at "ethics-washing" than substantive change in the industry.

As a result, much effort has been spent on ensuring the minimisation of bias in data, or on the presentation of data results in a manner that takes care of the data. We can, therefore, summarise this way of thinking through the notion of "bad data in, bad data out", or, as commonly understood in technology circles, "garbage in, garbage out". As a result, this often means that it is generally difficult for a user to verify, or question the results that computers generate, even as we increasingly rely on them for facts, news, and information. This confusion affects our understanding of not just an individual computer or software package, but also when the results are generated by networks of computers, and networks of networks. Thus, the "black box" is compounded into an illegible network, a system of opacity that, nonetheless, increasingly regulates and maintains everyday life, the economy, and media systems of the contemporary milieu. This has resulted in a number of technical challenges and responses by the programming industries.

Firstly, there has been an attempt to intimately link computation to the user through real-time computed results painted onto their screens. Computers and smartphones are not just information providers, but also increasingly also windows into marketplaces for purchasing goods, newspapers and magazines, entertainment centres, maps and personal assistants, etc.

This has intensified the intimate relationship between ourselves and *our* devices, *our* screens, *our* networks. But this creates its own problems, as the way in which the personal interface of the smartphone or computer flattens the informational landscape, and also has the potential for confusion between different functions and information sources, leads to post-truth claims and the derangement of knowledge.

Secondly, there is a temptation for the makers of these automated decision systems to use the calculative power of the device to persuade people to do things — whether buying a new bottle of wine, selecting a film to watch, or voting in a referendum or election. While the contribution of data science, marketing data, and persuasive technologies to the Trump election and the Brexit referendum remains to be fully explicated,[20] on a more mundane level, computers are active in shaping the way we think. The most obvious examples of this are Google Autocomplete on the search bar, which attempts to predict what we are searching for, and "infinite scroll" on social media networks and webpages, which are designed to capture attention and hold us trapped in their systems. Indeed, similar techniques have been incorporated into many aspects of computer interfaces, through design practices that persuade or nudge particular behavioural outcomes.

Thirdly, the large quantity of data collected, and the ease with which it is amassed and combined within new systems of computation, means that new forms of surveillance are beginning to emerge that go relatively unchecked. When this is combined with their seductive predictive abilities, real potentials for misuse or mistakes are magnified. For example, in Kortrijk, Belgium, and Marbella, Spain, the local police have deployed "body recognition" technology to track individuals by recognising their walking style or clothing; and across the European Union at least ten countries have a police force that uses face recognition.[21] Even with 99% accuracy in face recognition systems, the number of images in police databases makes false positives inevitable. Indeed, a 1% error rate means that 100 people will be flagged as wanted out of 10,000 innocent citizens. In The Netherlands, the police have access to a database of pictures of 1.3 million persons, many of whom have never been charged with a crime; in France, the national police can match CCTV footage against a file of 8 million people; and in Hungary, a recent law allows police to use face recognition in ID checks.[22] The lack of transparency in these systems, and the algorithms they use, is of growing social concern.

Fourthly, we see the emergence of systems of intelligence through technologies of machine learning and artificial intelligence. These systems do not only automate production and distribution processes, but also have the capacity to also automate consumption. The full implications of this are not just to proletarianise labour, but to proletarianise the cognitive abilities of people in society. This has made many formerly white-collar jobs redundant, but also serves to undermine and overtake the human faculty of reason. We see this in the creation of vast vertical and horizontal software infrastructures, which I have explored elsewhere through the notion of *infrasomatization* — the creation of cognitive infrastructures that automate value-chains, cognitive labour, networks, and logistics into new highly profitable assemblages built on intensive data capture.[23]

These technologies use the mobilisation of processes of selecting and directing activity, often through the automation of pattern matching, stereotypes, clichés, and simple queries.[24] But the underlying processes that calculate the results, and the explanation of how it was done, are hidden from the user — whether they are, for example, denied bail;[25] a loan;[26] insurance cover; or welfare benefits.[27] This explanatory deficit is a growing problem in our societies as the reliance on algorithms — some poorly programmed — creates potential situations that are inequitable and unfair, but also with little means of redress for citizens. This will be a growing source of discontent in society, but also may serve to delegitimate political and administrative systems which will appear as increasingly remote, unchecked, and inexplicable to members of society. Institutions and societies that rely heavily on these systems might then begin to suffer from a legitimacy crisis, as they are unable to change in response to social, political, and economic pressures, even as they generate socially unacceptable outcomes. Therefore, the capacity for explanatory thought, to ask the "why" questions of the computational systems that undergird and structure contemporary societies, becomes increasingly important.

Understanding the way in which the computational otherwise generates and magnifies uncertainty and a feeling of rising social risk and instability is also, to my mind, connected to a social desire for tethering knowledge, of grounding it in some way. We see tendencies generated by the liquidation of infor-mation modalities in "fake news", conspiracy theories, social media virality, and a rising distrust towards science and expertise, and the rise of relativism. This is also to be connected to new forms of nationalism, populism, and the turn to traditional knowledge and technical fixes to provide a new, albeit misplaced, ground for social epistemology. This new search for ground or foundations, whether through identity, tradition, formalism, or metaphysics is, to my mind, symptomatic of the difficulty of understanding and connecting computation and its effects across scales of individual and social life. As a result of this, computation itself becomes depoliticised and removed from public debate as a matter of concern — computation, then, seems to be merely technical, outside of political critique and, therefore, change. I also think we need to link this to the temptation for explanations that develop new metaphysics of the computational, which rely on formalism and mathematical axioms as an attempt to understand computation. These, I argue, seem to mirror the unhappy consciousness of neo-computationalism through a denial of the material in favour of a new form of idealism, allowing the actual existing political economy of computational capitalism to be ignored.

The challenge of new forms of social obscurity from the implementation of technical systems is given by the example of the machine-learning systems that have emerged in the past decade. As a result, a new explanatory demand has crystallised in an important critique of computational opaqueness and new forms of technical transparency. We see this, for example, in calls to ban facial recognition systems, public unease with algorithmic judicial systems, and with the passing of the California Consumer Privacy Act 2018 (CCPA).[28] Creel has usefully identified "functional, structural, or run transparency" as ways of thinking about explanation, but, here, I also want to add the importance of a social *right to explanation.* This has come to be identified as *explainability* within the fields of artificial intelligence and machine-learning, and requires a computational system that can provide an explanation for a decision it has made.

The *European Union General Data Protection Regulation 2016/679,* known as the GDPR, is key to helping us to understand this. This regulation creates the right "to obtain an explanation of [a] decision reached after such assessment and to challenge the decision" (GDPR 2016, Goodman et al. 2016). The GDPR creates a new kind of subject, the "data subject", to whom a right to explanation (amongst other data protection and privacy rights) is given. Additionally, it has created a legal definition of processing through a computer algorithm (GDPR 2016 Art. 4). Consequently, this has given rise to a notion of explainability which creates the right "to obtain an explanation of [a] decision reached after such assessment and to challenge the decision" (GDPR 2016 Recital 71). When instantiated in national legislation, such as the *Data Protection Act 2018* in the UK, a legal regime is created that can enforce a set of rights associated with computational systems. Important though this is, I argue that explainability is not just an issue of legal rights; it has also created a normative potential for a social right to explanation. The concept of explainability can be mobilised to challenge algorithms and their social norms and hierarchies, and it has the potential to contest platforms and automated decision systems. In relation to explanation, therefore, explainability needs to provide an answer to the question *why?* to close the gap in understanding. This raises a new potential for critique.

I now want to turn to thinking about what counts as an explanation, and how that might be related to computational systems. Hempel and Oppenheim argue that an explanation seeks to "exhibit and to clarify in a more rigorous manner" with reference to general laws. Some of the examples they give include a mercury thermometer, which can be explained using the physical properties of glass and mercury. Similarly, they present the example of an observer of a rowing boat, where part of the oar is submerged under water and appears to be bent upwards.[30] Under this definition, an explanation attempts to explain with reference to general laws. As Mill argues, "an individual fact is said to be explained by pointing out its cause, that is, by stating the law or laws of causation, of which its production is an instance".[31] Similarly, Ducasse argued in 1925 that "explanation essentially consists in the offering of a hypothesis of fact, standing to the fact to be explained as case of antecedent to case of consequent of some already known law of connection". Hempel and Oppenheim further argue that an explanation can be divided into two constituent parts, the *explanadum* and the *explanans*.[33] The explanandum is a logical consequence of the explanans. The explanans itself must have empirical context, which creates conditions for its testability. In this causal sense of explanation, science is often supposed to be the best means of generating explanations.[34] Explanations are assumed to tell us how things work, thereby giving us the power to change our environment in order to meet our own ends.

However, a causal mode of explanation is considered inadequate in fields concerned with purposive behaviour, as with computational systems, where the goals sought by the system are required in order to provide an explanation.[35] In this case, it might be more useful to ask: How long did the explanation take? Was it interrupted at any point? Who gave it? When? Where? What were the exact words used in the explanation? For whose benefit was it given? Indeed, it can be important to ascertain who created the explanation originally? Is it very complicated? In what form or medium of communication was it given?[36] For example, even after extensive discussion in the automotive industry about the ethics of driverless cars, Mercedes Benz has proposed that, in future, its own self-driving vehicles will be programmed to save the driver and the car's occupants in every situation, even if more pedestrians or road-users might be killed.[37] Requiring an explanation of such a system, and how it calculates the value of the lives of those affected, might be a key starting point for challenging the legitimacy of the assumptions built into it.

As a consequence, in understanding computational systems' reference to a teleological mode, rather than a causal mode of explanation has become more common. This approach has come to be called "machine behaviour", and tries to understand the "motivations" that guide the computational system. That is, to identify the goals sought by the system in order to provide an explanation.[38] Teleological approaches have the advantage that they make us feel that we really understand a phenomenon, because it describes things in terms of purposes with which we are familiar from our own experience of human goal-oriented behaviour. You can, therefore, see a great temptation to use teleological explanation in relation to AI systems, particularly by creating a sense of an empathetic understanding of the "personalities of the agents". But, both the causal and the teleological modes of explanation tend to create what we can think of as an "explanatory product". By explanatory product I mean that the outcome of an explanatory query might be a high-level diagram, technical description or list of counterfactuals, rather than any substantive explanation as to why a decision has been made. This has a number of limitations, including its static quality, and it might be more helpful when thinking about its potential for explanatory publics, to require a dynamic representation of the algorithmic process — how things were done, how they were computed. Many current discussions of explainability tend, chiefly, to be interested in an explanatory product, whereas I argue that an understanding of the explanatory process will have a greater impact on democratic politics. I would also like to connect this to the idea that an explanatory public might be able to "walk" through a contentious algorithmic decision by following the steps in a process, in order to understand how a decision was made. This would create the potential for discussing whether a decision was acceptable, allowing a public to understand how a decision was made and challenge the normative assumptions behind it.

Crucially, this connection between an explanatory product and the legal regime that enforces it has forced system designers and programmers to look for explanatory models that are sufficient to provide legal cover, but also at a level at which they are presentable to the user or data subject. But it remains uncertain if the "right is only to a general explanation of the model of the system as a whole ('model-based' explanation), or an explanation of how a decision was made based on that particular data subject's particular facts ('subject-based' explanation)".[39] This is not an easy requirement for any technical system, particularly in light of the growth of complicated systems of systems, and the difficulty of translating technical concepts into everyday language. It might, therefore, be helpful to think in terms of full and partial explanation, whereby a partial explanation is a final explanation with some part left out. That is, that in presenting a complicated system of automated decision systems, pragmatically, it is likely that explanations will assume an explanatory gap — assuming that the data subject is in possession of facts that do not need to be repeated. This, of course, may lead to the temptation to create persuasive, rather than transparent, explanations or a "good enough" explanation. This hints at the idea that those responsible for designing and building explainable systems will assume an underlying theory of general explainability and a theory of the human mind. These two theories are rarely explicitly articulated in the literature, and we need to better understand how they are deployed in explainable systems.

In conclusion, I have introduced and argued for the potential of a concept of explainability for developing a critique of the historically specific form of capitalist computation. In doing so, explainability and explanation can then be used to understand the ways in which justification and legitimacy are mobilised in computational societies. We must continually remind ourselves that the current information economy is historical. It owes its success and profitability to a legislative assignment of intellectual property rights and the amassing of data monopolies by the automatic operation of computers.[40] Other computations are possible, and different assemblages of computation and law might generate economic alternatives that mitigate or remove the current negative disruptive effects of computation in society. It is crucial to recognise that there is no "pure" or metaphysical computation, no privileged reading or access to an axiomatic or ontological computation — this identity thinking would be an objectivism which takes a single scientific or philosophical frame of reference as a given.[41] Rather, we must understand computation not so much in terms of an arbitrary attempt to make a metaphysics out of computation, but rather through its significance for us today. I argue that an encounter with computation takes different forms as history moves on. I have tried to show how we might do this through the mobilisation of concepts such as explainability, so that the underlying hylomorphism of computation may be understood. Through this the contradictions of computational capitalism might be laid manifest, and, more importantly, democratically challenged and potentially changed. I argue that this suggests that a rethinking of computation is needed in order to move it away from its current tendencies, from what I have called neo-computationalism, or *right computationalism,* which is geared towards some of the worst excesses of capitalism, and instead rethought within a new conception of *left computationalism.*[42] This would need to be developed through education and the capacity-building of explanatory publics, and by using critical concepts such as explainability to create the conditions for greater democratic thought and practice in computational capitalism.

**Notes**

1. See, for example, Jürgen Habermas, *The Structural Transformation of The Public Sphere* (Cambridge: Polity, 1992 [1962]); Seyla Benhabib, *Situating the Self: Gender, Community and Postmodernism in Contemporary Ethics* (London: Routledge, 1992); and Oskar Negt and Alexander Kluge, *Public Sphere and Experience: Toward an Analysis of the Bourgeois and Proletarian Public Sphere* (Minneapolis: University of Minnesota Press, 1993).

2. R. W. Outhwaite, "Laws and Explanations in Sociology", in *Classic Disputes in Sociology,* ed. R. J. Anderson et al. (Crows Nest, Australia: Allen & Unwin, 1987), 158.

3. Shane Richmond, "How Google crossed the creepy line", *The Telegraph,* 25 October 2010, https://www.telegraph.co.uk/technology/google/8086191/How-Google-crossed-the-creepy-line.html.

4. Sergey Brin and Larry Page, "The Anatomy of a Large-Scale Hypertextual Web Search Engine", in: *Seventh International World-Wide Web Conference* (WWW 1998), Brisbane, Australia, 14–18 April 1998.

5. Nicholas Carr, "Larry and Sergey: a valediction", *Rough Type,* 8 December 2019, http://www.roughtype.com/?p=8661%0D.

6. Peter Thiel, *Zero to One: Notes on Startups, or How to Build the Future* (New York: Crown Business, 2014).

7. Peter Thiel, "The Education of a Libertarian", *Cato Unbound,* 13 April 2009, https://www.cato-unbound.org/2009/04/13/peter-thiel/education-libertarian

8. Carr, "Larry and Sergey". See also David M. Berry, *Critical Theory and the Digital* (London: Bloomsbury. 2015), 178–182.

9. Steve Jobs, "When We Invented the Personal Computer…", *Computers and People Magazine,* July–August 1981, 8–11 & 22.

10. See David M. Berry, "The Contestation of Code: A Preliminary Investigation into the Discourse of the Free Software and Open Software Movement", in *Critical Discourse Studies,* 1 (1), January 2004, 65–89; and *Copy, rip, burn: the politics of copyleft and open source* (London: Pluto Press, 2008).

11. See Natasha Dow Schüll, *Addiction by Design: Machine Gambling in Las Vegas* (Princeton, NJ: Princeton University Press, 2014); and Nir Eyal, *Hooked: How to Build Habit-Forming Products* (London: Portfolio Penguin, 2014).

12. Bernard Stiegler, *The Automatic Society* (Cambridge: Polity Press, 2016); and Shoshana Zuboff, *The Age of Surveillance Capitalism: The Fight for a Human Future at the New Frontier of Power* (London: Profile Books, 2019).

13. David M. Berry, "Digital Breadcrumbs", *STUNLAW,* 14 October 2013, http://stunlaw.blogspot.com/2013/10/digital-breadcrumbs.html.

14. Matt Sledge, "CIA's Gus Hunt On Big Data: We 'Try To Collect Everything And Hang On To It Forever", *Huffington Post,* 20 March 2013, http://www.huffingtonpost.com/2013/03/20/cia-gus-hunt-big-data_n_2917842.html.

15. When accessing the *Huffington Post* on 16 January 2020, the following 29 companies were listed as collecting data: IAB partners, Active Agent AG, AdButler, Amazon, Atlas/Facebook, dataXtrade GmbH, DoubleClick/Google/Adx* (and its partners), eBay, Ensighten, Forensiq LLC, Mediametrie, Metrixlab, Microsoft, Nielsen Marketing Cloud, Otto (GmbH & Co KG), Paysafe/Income, Access, Pixalate, Inc., Plexop, Plexop (twice), PopWallet, Qriously, Quotient, salesforce.com, inc., Tchibo, Unruly Group Ltd, White Ops, Inc., Zentrick, zeotap GmbH. When one clicks through to the Double-Click Google partners, a further 163 data collection companies were revealed: 2KDirect Inc., PubMatic, hbfsTech, Rubicon Project, Amazon, Bannerflow, LiveRamp, Rakuten Marketing, Bucksense, Adara Media, IPONWEB, Weborama, Turbo, Jivox, Adform, Neustar, gskinner, Scenestealer, emetriq, Yieldr, mediasmart, Zebestof, MBR Targeting Gmbh, Adloox, Awin, Lotame usemax (Emego GmbH), Digilant, Avocet, The Reach Group, Integral Ad Science, UpRival, MiQ, ADman Media, MediaMath, Adverline, Sizmek, AdMaxim, Conversant/CJ Affiliate, DMA Institute, comScore, Kochava, Exactag, Widespace, Beeswax, affilinet, Akamai, Appreciate, Cloudflare, Neuralone, Innovid, FUSIO BY S4M, Meetrics, AppNexus, Neodata Group, nugg.ad, Commanders Act, Taboola, Cablato, Facebook, Visarity, Knorex, Nielsen, LifeStreet, Salesforce DMP, Adobe Advertising Cloud, Bombora, Scoota, OpenX Technologies, LKQD, Tradelab, AdLedge, Placecast, Lucid, Tealium, Roq.ad, Index Exchange, Piximedia, AdKernel, Platform161, Realzeit, TimeOne, Signal, Quantcast, Bidtheatre, Improve Digital, Pixalate, Teads, Smaato, Crimtan, Criteo, NEORY GmbH, DataXu, Zentrick, PulsePoint, advanced, STORE GmbH, TripleLift, Delta Projects, Demandbase, Fyber, Adludio, GlobalWebIndex, Arrivalist, Digitize, Media. net, DoubleVerify, ADEX, Simpli.fi, eBay, Centro, GroundTruth, Smart, smartclip Holding AG, Sharethrough, Inc., GetIntent, Sociomantic, Sojern, Eulerian Technologies, LoopMe, YieldMo, Impact, NEXD, Cuebiq, DYNADMIC, SpotX, Ligatus, Semasio GmbH, AdClear, Flashtalking, MainADV, Kantar, Mediarithmics, Virtual Minds, FreeWheel, Oath, travel audience — An Amadeus Company, Oracle Data Cloud, GroupM, Celtra, Publicis Media, Gemius, The Trade Desk, AudienceProject, AerServ, Adventori, Jampp, Videology, Captify, Google, White Ops, Tradedoubler AB, SpringServe,

Admetrics, Adacado, Exponential, Amobee, TrustArc, RTB House, Sublime Skinz, MGID, Underdog Media, Innity, Remerge. Note, these companies were deliberately listed out of alphabetical order making them even more difficult to follow and understand. For any consumer trying to get to grips with the data collection and spying on their behaviour, this list is impossible to control and manage and designed to work that way. The average website now has a large number of trackers, web bugs, beacons, and capture systems silently operating on their webpages; and the evidence points towards hidden tracking that is probably even worse on smartphone apps, which can more easily control, and access user data, sensor inputs, and screen use than browsers. See David M. Berry, "The Social Epistemologies of Software", *Social Epistemology,* Vol. 26, Nos. 3–4, October 2012, 379–398.

16. See Berry, *Critical Theory and the Digital,* 2015; Melissa Hellmann, "Special sunglasses, license-plate dresses: How to be anonymous in the age of surveillance", The Seattle Times, 12 January 2020, https://www.seattletimes.com/business/technology/special-sunglasses-license-plate-dresses-juggalo-face-paint-how-to-be-anonymous-in-the-age-of-surveillance; and Ramsey McGlazer, "Confessions of a Cake Boy", *Los Angeles Review of Books,* 6 January 2020, https://lareviewofbooks.org/article/confessions-of-a-cake-boy.

17. Gottfried Benn quoted in Peter Sloterdijk, *Critique of Cynical Reason,* trans. Andreas Huyssen (Minneapolis: University of Minnesota Press, 2010), 7.

18. Emily Dickinson, "The Lightening is a Yellow Fork", (n.d.).

19. Jane Wakefield, "Google's ethics board shut down", *BBC News,* 5 April 2019, https://www.bbc.co.uk/news/technology-47825833; and Sam

Shead, "Facebook Reportedly Has A Dedicated AI Ethics Team", *Forbes,* 3 May 2018, https://www.forbes.com/sites/samshead/2018/05/03/facebook-reportedly-has-a-dedicated-ai-ethics-team.

20. See Carole Cadwalladr, "Fresh Cambridge Analytica leak 'shows global manipulation is out of control'", *The Guardian,* 4 January 2020, https://www.theguardian.com/uk-news/2020/jan/04/cambridge-analytica-data-leak-global-election-manipulation.

21. Nicolas Kayser-Bril, "At least 10 police forces use face recognition in the EU, Algorithm-Watch reveals", *AlgorithmWatch,* 11 December 2019, https://algorithmwatch.org/en/story/face-recognition-police-europe.

22. Kayser-Bril, "At least 10 police forces use face recognition in the EU".

23. See David M. Berry, "Against infrasomatization: towards a critical theory of algorithms", in *Data politics: worlds, subjects, rights,* ed. Didier Bigo et al. (London: Routledge Studies in International Political Sociology, 2019), 43–63.

24. Catherine Malabou, *Morphing Intelligence: From IQ Measurement to Artificial Brains* (New York: Columbia University Press, 2019), 52; and Safiya Umoja Noble, Algorithms of Oppression (New York: New York University Press, 2018), 50.

25. Julia Dressel and Hany Farid, "The accuracy, fairness, and limits of predicting recidivism", *Science Advances,* Vol. 4, No. 1, DOI: 10.1126/sciadv.aao5580.

26. Amir Khandani et al*.,* "Consumer credit-risk models via machine-learning algorithms", *Journal of Banking & Finance,* Vol. 34, Issue 11, 2010, 2767–2787.

27. See also Virginia Eubanks, *Automating Inequality: How High-Tech Tools Profile, Police, and Punish the Poor* (New York, St. Martin's Press, 2018).

28. Natasha Singer, "What Does California's New Data Privacy Law Mean? Nobody Agrees", *New York Times,* 29 December 2019, https://www.nytimes.com/2019/12/29/technology/california-privacy-law.html.

29. Kathleen A. Creel, "Transparency in Complex Computational Systems", Philosophy of Science, Vol. 87, No. 4, October 2020, https://www.journals.uchicago.edu/toc/phos/2020/87/4.

30. Carl G. Hempel and Paul Oppenheim, "Studies in the Logic of Explanation", *Theories of Explanation,* edited by Joseph C. Pitt (Oxford: Oxford University Press, 1988), 9–50: 10.

31. John Stuart Mill, "Of the Explanation of the Laws of Nature", in *A System of Logic* (New York: Harpers & Brothers, 1858), 271–276.

32. C. J. Ducasse, "Explanation, Mechanism and Teleology 1", in *Truth, Knowledge and Causation* (London: Routledge, 2015 [1968]), 37.

33. Hempel and Oppenheim, "Studies in the Logic of Explanation", 10.

34. Joseph C. Pitt, ed., *Theories of Explanation* (Oxford: Oxford University Press, 1988), 7.

35. David-Hillel Ruben, *Explaining Explanation* (London: Routledge, 2016).

36. Ruben, *Explaining Explanation,* 6.

37. Charlie Sorrel, "Self-Driving Mercedes Will Be Programmed To Sacrifice Pedestrians To Save The Driver", *Fast Company,* 13 October 2016, https://www.fastcompany.com/3064539/self-driving-mercedes-will-be-programmed-to-sacrifice-pedestrians-to-save-the-driver.

38. See Dana Mackenzie and Judea Pearl, *The Book of Why* (London: Penguin Books, 2019), 364–365.

39. Lilian Edwards and Michael Veale, "Enslaving the Algorithm: From a 'Right to an Explanation' to a 'Right to Better Decisions'?" *IEEE*

*Security & Privacy,* Vol. 16, No. 3, July 2018, 46–54: 49.

40. David M. Berry, *Copy, Rip, Burn: The Politics of Copyleft and Open Source* (London: Pluto Press, 2008).

41. With the complexities raised by understanding computation and the new "ways of seeing" it makes possible, there is a strong temptation to ontologise computational categories and concepts. "Scholarship" may then turn to becoming further commentaries on the commentaries of these metaphysical claims, more conjectures on the conjectures. This is similar to what Adorno called identity thinking. It is therefore important to keep in mind that computational theory or mathematics does not "prove" what is metaphysically presupposed, or allow you to arbitrarily make a metaphysics out of mathematics or computation. Indeed, I would argue that we should avoid a formalist a priorism when attempting to understand a computational milieu. For more on this, see Berry, *Critical Theory and the Digital,* 2015.

42. By left computationalism and right computationalism, I am gesturing towards left and right Hegelianism.

### References

Anderson, R. J., Hughes, J. A. and Sharrock, W. W. eds. *Classic Disputes in Sociology.* Crows Nest, Australia: Allen & Unwin, 1987.

Benhabib, Seyla. *Situating the Self: Gender, Community and Postmodernism in Contemporary Ethics.* London: Routledge, 1992.

Berry, David M. "The Social Epistemologies of Software", *Social Epistemology,* Vol. 26, Nos. 3–4, October 2012, 379–398.

Berry, David M. "The Contestation of Code: A Preliminary Investigation into the Discourse of the Free Software and Open Software Movement". *Critical Discourse Studies,* Vol. 1, Issue 1 (January 2004): 65–89.

Berry, David M. *Copy, Rip, Burn: The Politics of Copyleft and Open Source.* London: Pluto Press, 2008.

Berry, David M. "Digital Breadcrumbs". *STUNLAW,* 14 October 2013. http://stunlaw.blogspot.com/2013/10/digital-breadcrumbs.html.

Berry, David M. *Critical Theory and the Digital.* London: Bloomsbury, 2015.

Berry, David M. "Against infrasomatization: towards a critical theory of algorithms". *In Data politics: worlds, subjects, rights,* edited by Didier Bigo, Engin F. Isin and Evelyn Ruppert. London: Routledge Studies in International Political Sociology, 2019. 43–63.

Brin, S. and Page, L. (1998) "The Anatomy of a Large-Scale Hypertextual Web Search Engine". In: *Seventh International World-Wide Web Conference* (WWW 1998), Brisbane, Australia, 14–18 April 1998.

Cadwalladr, C. "Fresh Cambridge Analytica leak 'shows global manipulation is out of control'". *The Guardian,* 4 January 2020. https://www.theguardian.com/uk-news/2020/jan/04/cambridge-analytica-data-leak-global-election-manipulation.

Carr, N. "Larry and Sergey: a valediction". *Rough Type,* 8 December 2019. http://www.roughtype.com/?p=8661%0D.

Creel, K. A. "Transparency in Complex Computational Systems". *Philosophy of Science,* Vol. 87, No. 4, October 2020. https://www.journals.uchicago.edu/toc/phos/2020/87/4.

Dressel, J. and Farid, H. "The accuracy, fairness, and limits of predicting recidivism". *Science Advances,* Vol. 4, No. 1. DOI: 10.1126/sciadv.aao5580.

Ducasse, C. J. "Explanation, Mechanism and Teleology 1". in *Truth,*

Knowledge and Causation. London: Routledge, 2015 [1968].

Edwards, L. and Veale, M. "Enslaving the Algorithm: From a 'Right to an Explanation' to a 'Right to Better Decisions'?" *IEEE Security & Privacy,* Vol. 16, No. 3, July 2018, 46–54.

Eubanks, V. *Automating Inequality: How High-Tech Tools Profile, Police, and Punish the Poor.* New York, St. Martin's Press, 2018.

Eyal, N. *Hooked: How to Build Habit-Forming Products.* London: Portfolio Penguin, 2014.

GDPR. "General Data Protection Regulation, Regulation (EU) 2016/679 of the European Parliament". 2016.

Goodman, B. and Flaxman, S. "European Union regulations on algorithmic decision-making and a 'right to explanation'". 2016. https://arxiv.org/abs/1606.08813.

Habermas, J. *The Structural Transformation of the Public Sphere.* Cambridge: Polity, 1992 [1962].

Hellmann, M. "Special sunglasses, license-plate dresses: How to be anonymous in the age of surveillance". *The Seattle Times,* 12 January 2020. https://www.seattletimes.com/business/technology/special-sunglasses-license-plate-dresses-juggalo-face-paint-how-to-be-anonymous-in-the-age-of-surveillance.

Jobs, S. "When We Invented the Personal Computer…" *Computers and People Magazine,* July–August 1981. 8–11 & 22.

Kayser-Bril, N. "At least 10 police forces use face recognition in the EU, AlgorithmWatch reveals". *AlgorithmWatch,* 11 December 2019. https://algorithmwatch.org/en/story/face-recognition-police-europe.

Khandani, A. E., Kim, A. J., and Lo, A. W. "Consumer credit-risk models via machine-learning algorithms". *Journal of Banking & Finance,* 34(11), 2010, 2767–2787.

Kuang, C. "Can A.I. Be Taught to Explain Itself?" *New York Times,* 21 November 2017 https://www.nytimes.com/2017/11/21/magazine/can-ai-be-taught-to-explain-itself.html.

Mackenzie, D. and Pearl, J. *The Book of Why.* London: Penguin Books, 2019.

Malabou, C. *Morphing Intelligence: From IQ Measurement to Artificial Brains.* New York: Columbia University Press, 2019.

McGlazer, R. "Confessions of a Cake Boy". *Los Angeles Review of Books,* 6 January 2020. https://lareviewofbooks.org/article/confessions-of-a-cake-boy.

Mill, J. S. *A System of Logic.* New York: Harpers & Brothers, 1858.

Negt, O. and Kluge, A. *A Public Sphere and Experience: Toward an Analysis of the Bourgeois and Proletarian Public Sphere.* Minneapolis: University of Minnesota Press, 1993.

Noble, S. U. *Algorithms of Oppression.* New York: New York University Press, 2018.

Pitt, J. C. ed. *Theories of Explanation.* Oxford: Oxford University Press, 1988.

Richmond, S. "How Google crossed the creepy line". The Telegraph, 25 October 2010. https://www.telegraph.co.uk/technology/google/8086191/How-Google-crossed-the-creepy-line.html.

Ruben, D.-H. *Explaining Explanation.* London: Routledge, 2016.

Sample, I. "Computer says no". *The Guardian,* 5 November 2017. https://www.theguardian.com/science/2017/nov/05/computer-says-no-why-making-ais-fair-accountable-and-transparent-is-crucial.

Schüll, N. D. *Addiction by Design: Machine Gambling in Las Vegas.* Princeton, NJ: Princeton University Press, 2014.

Shead, S. "Facebook Reportedly Has A Dedicated AI Ethics Team". *Forbes,* 3 May 2018.

https://www.forbes.com/sites/samshead/2018/05/03/facebook-reportedly-has-a-dedicated-ai-ethics-team.

Singer, N. "What Does California's New Data Privacy Law Mean? Nobody Agrees". *New York Times,* 29 December 2019. https://www.nytimes.com/2019/12/29/technology/california-privacy-law.html.

Sledge, M. "CIA's Gus Hunt On Big Data: We 'Try To Collect Everything And Hang On To It Forever". *Huffington Post,* 20 March 2013. http://www.huffingtonpost.com/2013/03/20/cia-gus-hunt-big-data_n_2917842.html.

Sloterdijk, P. *Critique of Cynical Reason.* Translated by Andreas Huyssen. Minneapolis: University of Minnesota Press, 2010.

Sorrel, C. "Self-Driving Mercedes Will Be Programmed To Sacrifice Pedestrians To Save The Driver". *Fast Company,* 13 October 2016. https://www.fastcompany.com/3064539/self-driving-mercedes-will-be-programmed-to-sacrifice-pedestrians-to-save-the-driver.

Stiegler, B. *The Automatic Society.* Cambridge: Polity Press, 2016.

Thiel, P. "The Education of a Libertarian". *Cato Unbound,* 13 April 2009. https://www.cato-unbound.org/2009/04/13/peter-thiel/education-libertarian.

Thiel, P. A. *Zero to One: Notes on Startups, or How to Build the Future.* New York: Crown Business, 2014.

Turek, M. "Explainable Artificial Intelligence (XAI)". *Defense Advanced Research Projects Agency,* (n.d.). https://www.darpa.mil/program/explainable-artificial-intelligence.

Wakefield, J. "Google's ethics board shut down". *BBC News,* 5 April 2019. https://www.bbc.co.uk/news/technology-47825833.

Zuboff, S. *The Age of Surveillance Capitalism: The Fight for a Human Future at the New Frontier of Power.* London: Profile Books, 2019.